\documentclass[
reprint,
 amsmath,amssymb,
 aps,
 prl
]{revtex4-1}

\usepackage{graphicx}% Include figure files
\usepackage{dcolumn}% Align table columns on decimal point
\usepackage{bm}% bold math
\newcommand{\tzimm}{\tau_{z}}

\newcommand{\DK}{D_\textrm{K}}
\newcommand{\kB}{k_\textrm{B}}
\newcommand{\tZ}{\tau_\textrm{Z}}
\newcommand{\AF}{A_\textrm{F}}

\begin{document}

\preprint{APS/123-QED}

\title{Time Reversal of the Overdamped Langevin Equation and Fixman's Law}% Force line breaks with \\

\author{Oliver T. Dyer}
\author{Robin C. Ball}%
 \email{r.c.ball@warwick.ac.uk}
\affiliation{%
 Dept of Physics, University of Warwick, Coventry CV4 7AL, UK.
}%


\date{\today}% It is always \today, today,

\begin{abstract}
%% don't take this too seriously.  For now it is just
%% me mapping out a storyline
We discuss how the first order Langevin equation for the overdamped dynamics of an interacting system has a natural time reversal of simple but surprising form, with consequences for correlation functions.  This leads to the correlation of interactions as a strictly restraining term in the time-dependent diffusion tensor of the system, deriving the relation first suggested by Fixman. Applying this to the time-dependent diffusion of dilute polymer coils leads to the quantitative calibration of Kirkwood's approximation for their hydrodynamic radius.  We find the generalized ``Fixman Law" for dissipation with a memory kernel, which has revealing causal structure, and we also discuss the case of the second order Langevin Equation.
 
%\begin{description}
%May be entered using the %\verb+\pacs{#1}+ command.
%You may use the %\texttt{description} environment %to structure your abstract;
%use the optional argument of the %\verb+\item+ command to give the %category of each item. 
%\end{description}
\end{abstract}

\pacs{Valid PACS appear here}% PACS, the Physics and Astronomy
\maketitle

%\tableofcontents

The first order Langevin equation  \cite{langevin1908} 
\begin{equation}\label{eq:LangevinNoMem}
\frac{dx_i}{dt}=M_{ij}f_j(x(t))+\kB T \partial_k M_{ki} + u_i(t) 
\end{equation}
describes the dynamics of a subsystem of coordinates $x_i(t)$ with  interaction forces $f_i(x)$, embedded in (or part of) a larger background system in thermal equilibrium at temperature $T$ with short timescale response. Being rooted in equilibrium it should have a time reversal symmetry, and we show how to find this and non-trivial  fluctuation theorem results which follow.  In later sections we show a notable application and discuss the generalisation of our results to both inertial memory and Langevin dynamics in memory media.

The background system being in thermal equilibrium contributes  random velocity contributions $u_i(t)$ with correlation  $\langle u_i(t)u_j(t') \rangle=2\kB T M_{ij}(x)\delta(t-t')$  related to the mobility $M_{ij}$ by the Fluctuation-Dissipation Theorem  \cite{kubo1957}.  We adopt It\^{o} calculus in taking the $u_i(t)$ to be uncorrelated with present as well as past positions, requiring us to include the explicit Brownian drift term $\kB T \partial_k M_{ki}$ \cite{laulubensky}.

We now seek a time reversal of Eq.~\eqref{eq:LangevinNoMem} with $t^R=-t$ and $x^R(t^R)=x(t)$ obeying
\begin{equation}\label{eq:TimeReverseLangevin}
\frac{dx_i^R}{dt^R}=M_{ij}f_j(x^R(t^R))+\kB T \partial_k M_{ki} +u_i^R(t^R)
\end{equation}
which requires that the time reversal of the random velocity contribution be given by
\begin{equation}
u_i^R(t^R)=-u_i(t)-2v_i(t) \label{eq:reversedrandomcurrent},
\end{equation}
where 
%%\begin{equation}\label{eq:videf}
$    v_i(t)=M_{ij}f_j(x(t)) + \kB T \partial_k M_{ki} $
%%\end{equation}
is the deterministic contribution to the forwards motion of the system.

At first sight the deterministic contribution to the time reverse random velocity is puzzling because the random velocity should be unbiased, but this is conditional on the direction of time:  $u(t)$ is uncorrelated with past and present $x(t'\le t)$.  Correspondingly $u^R(t^R)$ should be uncorrelated with future and present $x(t'\ge t)$, and the difference of condition matters because $x(t'\ge t)$ clearly cumulates influence from the earlier random velocity terms $u(t)$. 

Useful correlation identities follow from Eq.~\eqref{eq:reversedrandomcurrent} by considering that for $t^R \ge t'^R$ we  have 
$\langle u_i^R (t^R)v_j^R (t'^R)\rangle=0$
where $v_i^R(t^R)=v_i(t)$.  Substituting back in terms of unreversed quantities then leads to
\begin{equation}
\langle u_i(t)v_j(t')\rangle = -2 \langle v_i(t)v_j(t') \rangle , \quad t<t'
\end{equation}
whereas $\langle u_i(t)v_j(t')\rangle =0$ for $t>t'$.  From this the full velocity autocorrelation of the interacting system   can be expressed as 
\begin{equation}
\langle \dot{x_i}(t)\dot{x_j}(t') \rangle=2M_{ij}/(\kB T)\,\delta(t-t')- \langle v_i(t)v_j(t')\rangle.\label{Eq:fixmanlaw}
\end{equation}
%%The above follows by substituting $dx_i/dt=v_i(t)+u_i(t)$ and using our relation further above to eliminate cross terms.

The power of Eq. (\ref{Eq:fixmanlaw}) is that it expresses the diffusion of the interacting system in terms of the bare non-interacting value \emph{minus} a restraining contribution from the deterministic motion. This result was first proposed by Fixman by Diffusion Equation arguments \cite{fixman1981}, in the polymer context (without Brownian drift) which we discuss below, but in a manner which did not convince later authors albeit they found some numerical evidence to support it when $t\simeq t'$ \cite{dunweg}.  As it is now proved, we will refer to it as \emph{Fixman's Law}.

\section*{Application to Polymer Diffusion}

The archetypal application is to the motion of colloid and polymer systems in a Newtonian solvent, where the \(x_i(t)\) are the \(3N\) coordinates of \(N\)particle position vectors \(\vec{r}_n(t)\) and radius $a_n$, with joint mobility tensor  given by \(O_{nn}=I/\left(6\pi \eta a_n\right)\) for diagonal blocks (with $I$ the three dimensional identity matrix) and essentially that of Oseen with \(M\rightarrow O_{mn}=\left( I+\hat{r}\hat{r}\right)/\left(8\pi \eta |\vec{r}_m-\vec{r}_n|\right) \) for off-diagonal blocks \(m\ne n\), but matching onto the diagonal case at close approach \cite{rotneprager,dyerball}. Being based on divergence-free Stokes flow, these mobility tensors have zero Brownian drift.    For an isolated polymer coil of \(N\) beads Eq.~\eqref{Eq:fixmanlaw} then becomes

\begin{equation}
\langle    \dot{ \vec{r}}_n(t)   \dot{ \vec{r}}_m(t')  \rangle =2 \kB T \, \langle O_{mn} \rangle \delta(t-t') -  \langle   \vec{v}_n(t)   \vec{v}_m(t')  \rangle 
\end{equation}
where the leading term is Kirkwood's approximation\cite{kirkwood1954} and all the time dependent memory is captured in the counter terms correlating the velocity contribution 
\( \vec{v}_m(t)=\sum_n  O_{mn}  \cdot  \vec{F}_n(t) \) 
driven by the conservative forces holding the polymer chain together.  

Focussing for simplicity on the coil centre of mass \(\vec{R}(t)=\sum_n \vec{r}_n(t)/N\), the corresponding velocity autocorrelation function is then given by 
\begin{equation}
 \langle    \vec{V}(t) \vec{V}(t')  \rangle = 2 \DK \delta(t-t')I - \AF(t-t')I
\end{equation}where \(\DK=\kB T N^{-2}\sum_{mn} \langle O_{mn} \rangle\) is the classical Kirkwood diffusivity\cite{kirkwood1954} and all the memory in the polymer centre of mass motion comes from the Fixman counter-term
\begin{equation}
\AF(t-t')I =N^{-2} \! \! \! \! \sum_{mnm'n'} \! \langle  O_{mm'}(t)\cdot \vec{F}_{m'}(t)   \vec{F}_{n'}(t')  \cdot O_{n'n}(t') \rangle  \label{Eq:AFfixman}.
\end{equation}
The long time polymer coil diffusivity is then given by
\begin{equation}D_L=\DK-\Delta=\DK-\int_0^\infty \AF(\tau)d\tau \label{eq:DL}
\end{equation}
and given that \(D_L\) is the natural experimental measurement whilst \(\DK\) is a direct configurational moment, it is important to understand their difference \(\Delta\), all arising from the Fixman term. 

\begin{figure}
\includegraphics[width=\linewidth]{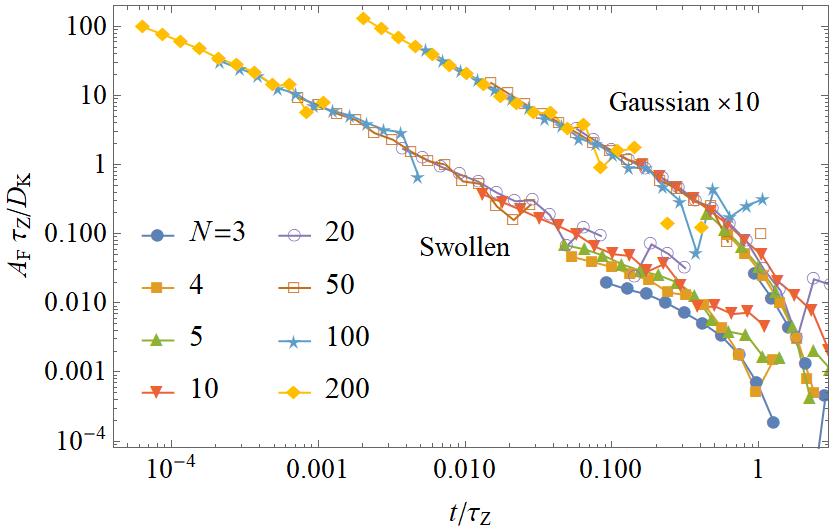}
\centering
\caption{Autocorrelation of the force driven Centre of Mass velocities $\AF(t)$ (as defined in Eq. \ref{Eq:AFfixman}) \emph{vs} time $t$ for swollen chains (lower curves) and for Gaussian chains (upper curves with $\AF$ shifted up by a factor of 10).  
With units naturally scaled in terms of measured Zimm time $\tZ$ and Kirkwood (short time) diffusion coefficient $\DK$, there is good superposition for the wide range of chain length $N$ shown.  The plots exclude measurements at times short in absolute (rather than scaled) units, for which each curve breaks away from the superposition (see text). 
}  
\label{Fig: CAA vs scaled time}
\end{figure}

\begin{figure*}
\centering
\includegraphics[width=1.0\columnwidth]{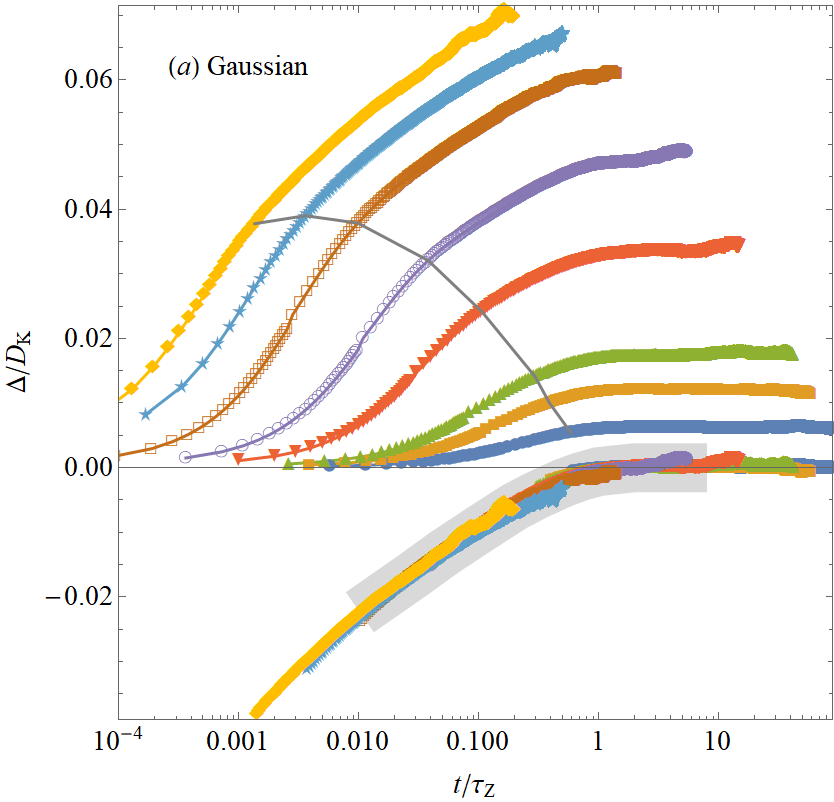}
\includegraphics[width=1.0\columnwidth]{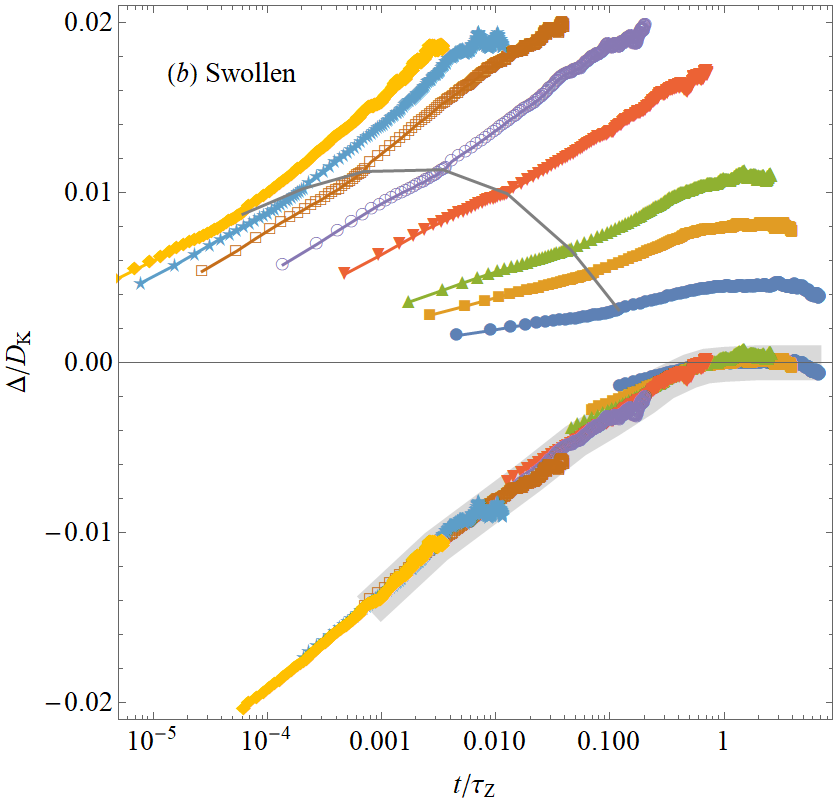}
\caption{Upper curves:  fractional decrease in diffusivity, found by numerically integrating the data in Fig.~\ref{Fig: CAA vs scaled time} for the autocorrelation function $\AF(t)$, plotted against the upper time limit of integration.  
Chain lengths are as per legend in Fig.~\ref{Fig: CAA vs scaled time}.
Data has been truncated when noise begins to dominate additional contributions leading to incomplete curves, especially for swollen chains. 
The lower curves show the same data shifted vertically to a common master curve, with plateau value zero and shown in gray, the shift for each curve then giving an estimate of the eventual long time plateau of the relative decrease in diffusivity.
The sections of each curve below the gray arcs in the upper plots are omitted to make the master curves clearer.
Fig.~\ref{Fig: dD/DK2} in the Supplementary Material gives a different presentation of these master curves.}
\label{Fig: dD/DK}
\end{figure*}

In Fig.~\ref{Fig: CAA vs scaled time} we show the time-dependence of $\AF(t)$ in Gaussian and swollen chains measured in Wavelet Monte Carlo dynamics simulations \cite{dyerball,dyerthesis}. In each case the longer time data is tolerably consistent with the natural scaling form  $\AF(t)=\DK/\tZ~ \mathfrak{h}(t/\tZ)$ as plotted where we have used the relaxation time  $\tZ$ of the longest Rouse mode as the natural scale of time, and this behaviour on its own would lead upon time integration of $\AF$ to a contribution to $\Delta$ proportional to $\DK$.   Outside the range of data shown in Fig.~\ref{Fig: CAA vs scaled time} we also see short time structure $\AF(t)=\mathfrak{g}(N) \mathfrak{f}(t)$ on monomeric timescales which on its own would lead to a contribution to $\Delta$ proportional to $\mathfrak{g}(N)$, and for the particular case $t=0$ earlier work \cite{dunweg} showed that $\mathfrak{g}(N)$ does appear to approach $1/N$ albeit slowly.
%, and recent work \cite{dyerthesis} obtaining  $\log(N)/N$ for the Gaussian chain case.

In Fig.~\ref{Fig: dD/DK} we show the corresponding cumulative integrals for $\Delta(t)=\int_0^t \AF(t')dt'$ in scaled variables, as far as we can measure them above noise. Assuming these plots do go far enough to capture all the short time parts, the total values of $\Delta_N/\DK$ inclusive of the scaling parts are estimated by finding the vertical shifts to align to a master curve in the scaled variables with plateau value zero.

\begin{figure}
\includegraphics[width=\linewidth]{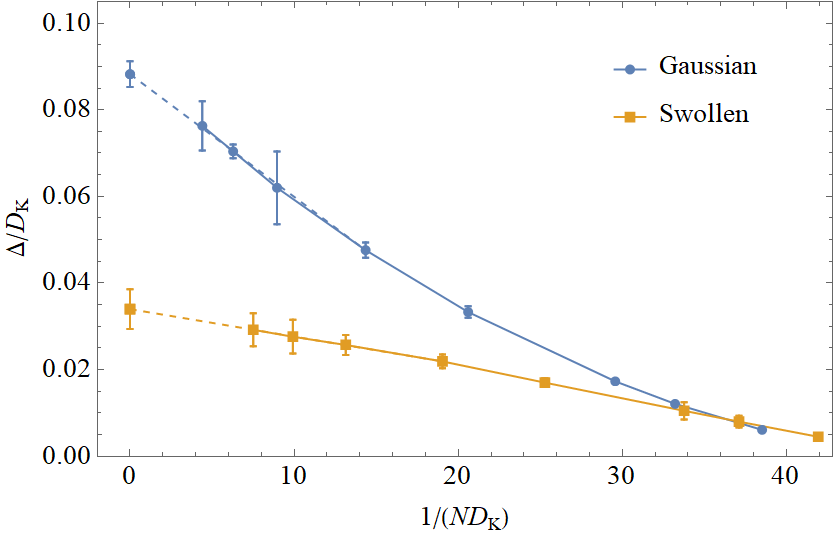} 
\centering
\caption{
Plateau values of $\Delta/\DK$ for different polymer chain lengths, plotted against the correction to scaling variable $1/(N\DK)$. 
Data for long chains were obtained using the master curves in Fig.~\ref{Fig: dD/DK} to extrapolate out to the plateau. 
The error bars on these data show the difference between this master curve extrapolation and the estimated upper bound from a linear extrapolation. 
For the data that did reach $t/\tZ = 1$, the error bars show twice the standard deviation of $\Delta/\DK$ measured across all simulations contributing to each data point. 
The dashed lines to the left of the data indicate fits of the 4 leftmost data points with the function $m/(N\DK)+c$, and the data at $1/(N\DK) = 0$ show the extrapolated infinite chain value with error bars calculated as $2\sqrt{\sigma_{cc}^2}$, with $\sigma^2$ the covariance matrix for $m$ and $c$.
}  
\label{Fig: extrapolateddDonD}
\end{figure}

The measured estimates of $\Delta_N/\DK$ are shown in Fig.~\ref{Fig: extrapolateddDonD} plotted against the anticipated correction to scaling which is $N^{-1}/R_H^{-1}$. These enable us to give the first quantitative estimates of the asymptotic values for true long time diffusion coefficient compared to the Kirkwood short time formula\cite{kirkwood1954},
\begin{equation} 
\lim_{N\to \infty}\frac{D_L-\DK}{\DK}= \begin{cases}  
(-3.4 \pm 0.5)\% & \textrm{swollen chains} \\
(-8.9 \pm 0.5)\% & \textrm{Gaussian chains}.  \label{eq:extrapolateddDonD}
\end{cases}
\end{equation}
These values \cite{rmk:errors} considerably firm up earlier ones, where for phantom chains Fixman \cite{fixman1986} concluded the deficit was \emph{around} 8\% whilst for swollen chains Liu \& D\"{u}nweg \cite{dunweg} concluded it was \emph{at least} 3.5\%.

These corrections matter wherever a measured chain diffusivity or mobility, such as might be observed by dynamic light scattering or sedimentation respectively, is compared with the Kirkwood formula given in terms of configurational statistics. Direct determination of these corrections to our accuracy from simulation without use of the Fixman law would be very hard, as this would require direct measurement of $D$ to better than $1\%$ which would entail simulation (with full hydrodynamics included) in excess of $10^4$ chain relaxation times for each chain length.  Direct determination by experiment would face the added difficulties of determining the configurational statistics to better than $1\%$ and controlling any influence of polydispersity.

\section{Second Order Langevin Equation}

The basis for the second order Langevin equation is slightly different, but it leads to a matching time reversal result.  Working in terms of forces we have 
\begin{equation}
m_{ij}\frac{d^2x_j}{dt^2}+Z_{ij}\frac{dx_j}{dt}=f_i(x(t))+\phi_i(t). \label{eq:Langevin2}
\end{equation}
with random forces \(\phi_i(t)=Z_{ij}u_j\)  where $Z=M^{-1}$ is the impedance matrix and $m$ an inertial tensor. 

We now pose the analogous time reversed equation
\begin{equation}
m_{ij}\frac{d^2x^R_j}{(dt^R)^2}+Z_{ij}\frac{dx^R_j}{dt^R}=f_i(x^R(t^R))+\phi^R_i(t^R),
\end{equation}
with \(x^R(t^R=-t)=x(t)\) as before, and by inspection this requires 
\begin{equation} \phi^R_i(t^R)=\phi_i(t)-2Z_{ij}\frac{dx_j}{dt}.
\end{equation}This looks quite different from the first order case, but it turns out to exactly agree in the limit of inertia being negligible.  Multiplying through by the mobility matrix to obtain the random currents, eliminating  \(Z\cdot dx/dt\) using the second  order Langevin equation \eqref{eq:Langevin2}, and then noting that $M_{ij} \ddot{x}_j \rightarrow -\kB T \partial_j M_{ji}$ in the limit for the first order Langevin equation leads back to the time reversal of the random currents as in Eq. \eqref{eq:reversedrandomcurrent}.

\section{Generalisation to Memory Media}

Finally we have considered the generalisation to the first order Langevin equation for motion in a time dependent (or memory) medium, seeking insight into the nature of time reversal, and status of Fixman's law in relation to other fluctuation-dissipation type results. The generalisation turns out to be quite distinctive, confirming that Fixman's law \eqref{Eq:fixmanlaw} is self-standing result and not a rewriting of prior fluctuation theorems.  We start from 
\begin{equation} \label{eq:memorylangevin}
    \frac{dx}{dt}=\int G(t-t')f(x(t'))dt'+u(t)=G*f+u
\end{equation}
where $G(t-t')$ is the response function of the medium which we now take to depend only on explicit time difference and not on $x(t)$, so there is no Brownian drift.  Causality dictates that  $G(t-t')=0$  for $ t<t'$,  and for the forces we will use the shorthand notation $f(t)=f(x(t))$.  We leave indices implicit now but for the multivariate case $G$ is a matrix, in terms of which the Fluctuation-Dissipation Theorem \cite{kubo1957} gives
\begin{equation}\left\langle u(t)u^{T}(t')\right\rangle =kT\left(G(t-t')+G^{T}(t'-t)\right).\end{equation}

We will allow the memory medium to be non-reversible, in particular in terms of its  off-diagonal elements of $\left\langle u(t)u^T(t')\right\rangle$. This then means that whilst we seek a time reversal transformation preserving the form of the Langevin Equation as a class, it will no longer in general be a symmetry for each case.  The transformation is found as before by negating all time arguments to $t^{R}=-t$ in our equation
of motion, leading to   \(
-dx^{R}/dt^{R}=G^{R}*f^{R}+u(-t^R)
\), 
where $x^{R}(t^{R})=x(t)=x(-t^{R})$ and similarly for $f^{R}$ and $G^{R}\equiv G^{R}(t^{R})=G(t)=G(-t^{R})$.  This matches the natural time reverse equation
\begin{equation}
  dx^{R}/dt^{R}={G^T}*f^{R}+u^R(t^R) 
\end{equation}
by taking
\begin{equation}
 u^R(t^R)= -u(-t^R)-(G^T+G^{R})*f^R . \label{eq:timereversalmemory} 
\end{equation}

This time reversal of the random velocities  \eqref{eq:timereversalmemory} features terms \(G^T*f^R\) which is causal in reversed time and  \(G^{R}*f^R\) which is anti-causal in their propagation of the conservative forces, complicating derivation of results from it. We   show in the supplementary material that it follows at some length that 
\begin{equation} \langle u f^T  \rangle = - (G+G^{TR})*\langle f f^T  \rangle_-\end{equation}
where  $\langle f(t) f^T(t') \rangle_{-,+}$ are respectively the $t<t'$ and $t>t'$ parts of $\langle f(t) f^T(t') \rangle$.
It is then a matter of straightforward substitution to find the generalisation of Eq.~\eqref{Eq:AFfixman} as
\begin{equation}\label{eq:AFfixmanMM}
\begin{split} 
\AF&=\left\langle \dot{x}\dot{x}^{T}\right\rangle _{\textrm{free}}-\left\langle \dot{x}\dot{x}^{T}\right\rangle \\&=G*\left\langle ff^{T}\right\rangle _{+}*G + G^{TR}*\left\langle ff^{T}\right\rangle _{-}*G^{TR}.
\end{split}\end{equation}

\begin{figure}
\includegraphics[width=\linewidth]{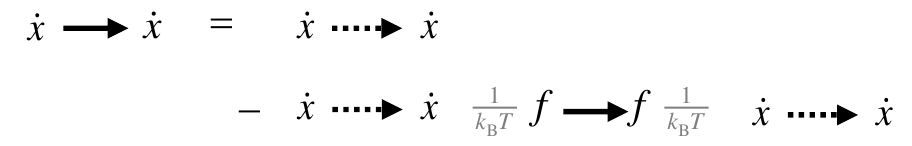}
\centering
\caption{
Diagrammatic representation of Fixman Law for a memory medium. Full lines denote the total correlation whilst dashed lines denote their correlations without interaction, as per the background medium.  All the correlations are consistently directed in time as shown by the arrows.}
\label{fig:diagram} 
\end{figure} 

This memory medium version of Fixman's law \eqref{eq:AFfixmanMM} is richer in structure than conventional Fluctuation-Dissipation Theorem type results. Specialising to \(t>t'\) for clarity, \(\AF(t-t')\) can be expressed as
\[\int\displaylimits_{t>u>u'>t'} \mkern-20.0mu du\,du'\left\langle G(t-u)\!\cdot\!f(u) \;   (G^R(t'-u')\!\cdot\!f(u'))^T\right\rangle
\]
which is the correlation of the causal response to the applied causal forces with  the corresponding anti-causal response, with the added restriction that only causally ordered factors of forces contribute. Substituting the $G$ in terms of  correlation functions the full result is represented diagrammatically in Fig. \ref{fig:diagram}, where the time ordering of its terms can be seen to be natural.  

Further specialisations, generalisations and applications await attention.  Without memory media Langevin equations describe a Markov process so everything can be worked through in terms of Fokker-Planck equations.  For the dynamics in memory media we have only considered their response depending on the time lag, whereas it is natural to consider complex fluid media with response dependent on the full configurational history. The application to colloid dynamics is interesting because Fixman's law separates the contribution of hydrodynamic and conservative interactions, whilst for polymers we would like to generalise to encompass rheological response functions.  

\begin{acknowledgments}
This work has been supported by the Engineering and
Physical Sciences Research Council (EPSRC), Grant No.s
EP/L505110/1 and EP/M508184/1 (PhD studentship for O.T.D.).
Computing facilities were provided by the Scientific Computing Research Technology Platform of the University of Warwick.
\end{acknowledgments}

\bibliography{ms_refs}

\begin{thebibliography}{12}
\bibitem [{rmk()}]{rmk:errors}%
  \BibitemOpen
  \href@noop {} {}\bibinfo {note} {We have rounded up the statistical error on
  the Gaussian limiting value, as it is based on rather scattered error bars in
  Fig. \ref{Fig: extrapolateddDonD}}\BibitemShut {NoStop}%
\end{thebibliography}

%\begin{thebibliography}{99}
%\bibitem{einstein1905}A. Einstein, Ann. der Physik {\bf 17} 549-560 (1905).
%\bibitem{langevin1908}P. Langevin, {\it Sur la théorie du mouvement brownien}, C. R. Acad. Sci. Paris {\bf 146} 530-533 (1908).

%\bibitem{laulubensky}A. W. C. Lau and T. C. Lubensky, Phys. Rev. E 76, 011123 – Published 27 July 2007.

%\bibitem{kubo1957}R. Kubo, J. Phys. Soc. Japan {\bf 12} 570-586 (1957).

%\bibitem{fixman1}M. Fixman, J. Chem. Phys., 45, 785 (1966).

%\bibitem{earliest}from hortafixman: (3) Y. Ikeda, Kobayashi Rigaku Kenkyusho Hokoku, 6, 44 (1956).
%(4) J. J. Erpenbeck and J. G. Kirkwood, J. Chem. Phys., 38, 1023 (1963).

%\bibitem{hortafixman}Horta A, Fixman M. Translational diffusion constant of polymer chains Journal of the American Chemical Society. 90: 3048-3055. Approximate estimates 1.4\% and lower.

%\bibitem{fixman1981}M. Fixman, Macromolecules {\bf 14}, 1710 (1981)

%\bibitem{rmk:errors}We have rounded up the statistical error on the Gaussian limiting value, as it is based on rather scattered error bars in Fig. \ref{Fig: extrapolateddDonD}

%\bibitem{fixman1986}M. Fixman, J. Chem. Phys. {\bf 84}, 4080 (1986)

%\bibitem{dunweg}B. Liu and B. D\"{u}nweg, J. Chem. Phys. {\bf 118}, 8061 (2003)

%\bibitem{dyerball} O. T. Dyer and R. C. Ball, J. Chem. Phys. {\bf 146}, 124111 (2017)

%\bibitem{dyerthesis} O. T. Dyer, PhD Thesis, University of Warwick, UK (2019)

%\bibitem{OUprocess} G. E. Uhlenbeck, L. S. Ornstein, Phys. Rev. {\bf 36}, 823-841 (1930).  doi:10.1103/PhysRev.36.823.

%\end{thebibliography}

\section{Supplementary Material}

\section{Fixman's Law in a Memory Medium}

\begin{figure*}
\centering
\includegraphics[width=\columnwidth]{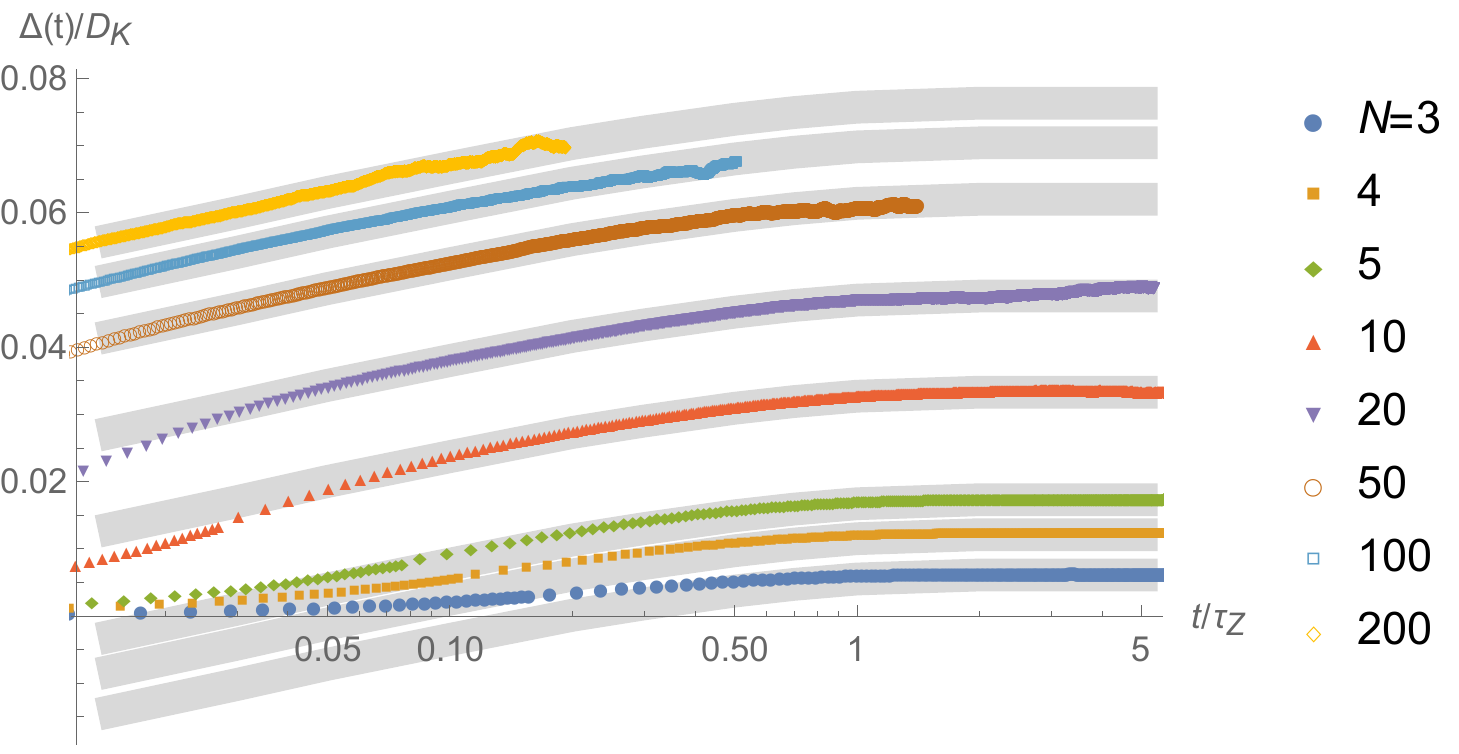}
\includegraphics[width=\columnwidth]{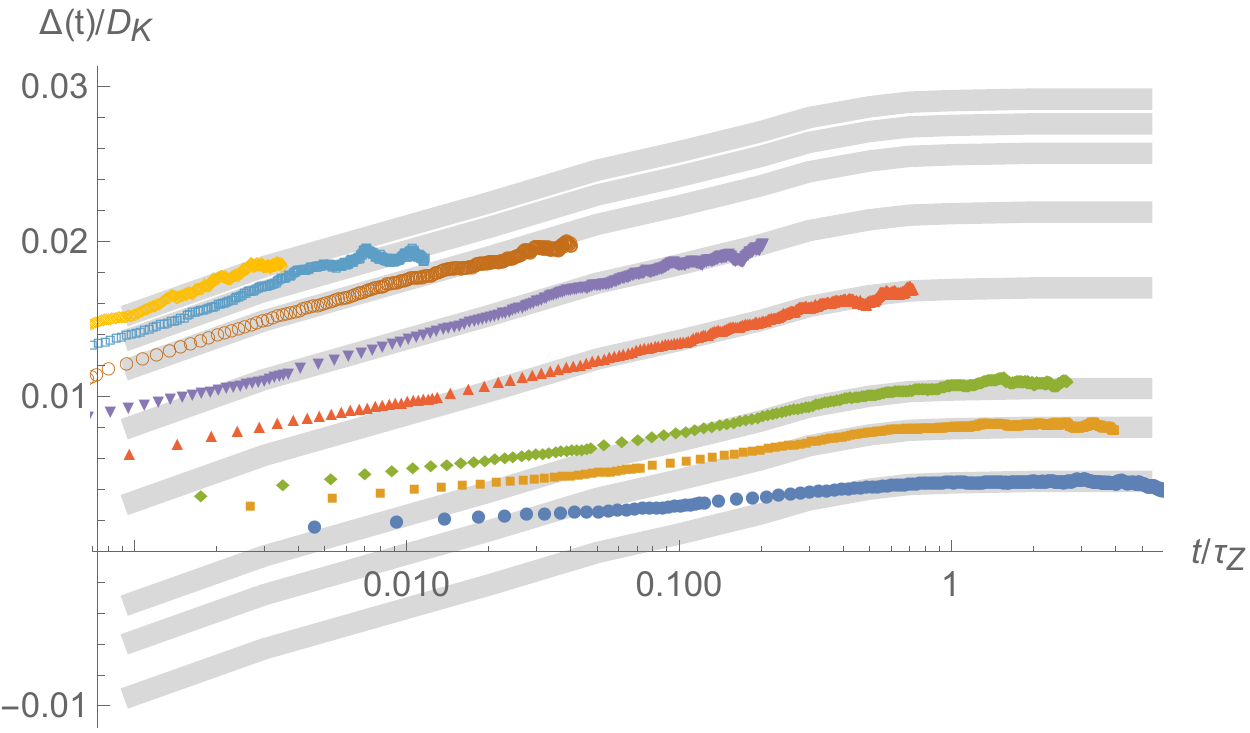}
\caption{Fractional decrease in diffusivity, found by numerically integrating the data in Fig.~\ref{Fig: CAA vs scaled time} for the autocrrelation function $\AF(t)$, plotted against the upper time limit of integration, for various chain lengths $N$. Master curves shown in gray were obtained by vertical linear shift of the data in the long time regime, and these are shown in gray behind (and extending) the curves for each $N$.  The shifted master curves then give an estimate of the eventual long time plateau of the relative decrease in diffusivity for each curve.}
\label{Fig: dD/DK2}
\end{figure*}

Our general equation of motion is
\begin{equation}
\frac{dx}{dt}=\int G(t-t')f(x(t'))dt'+u(t)
\end{equation}
where for causality $G(t-t')=0,\quad t<t'$ and we will also allow
ourselves the shorthand notation $f(t)=f(x(t))$. For the multivariate
case, that is where $x$, $f$ and $u$ all become time dependent
vectors, the above displayed equation applies with $G$ being a matrix and we have
$\left\langle u(t)u^{T}(t')\right\rangle =\kB T\left(G(t-t')+G^{T}(t'-t)\right)$. 

It proves useful to express the random velocity contributions as 
\begin{equation}
u(t)=\int K(t-t')h(t')dt'
\end{equation}
where $\left\langle h(t)h^{T}(t')\right\rangle =\delta(t-t') I$
is white noise and $K(t-t')=0$ for $t<t'$ is strictly causal. Then
we can interpret $h(t)$ as the innovation, that is what is new in
the noise at time $t$, with the important consequence that $h(t)$
is uncorrelated with everything from earlier times, including

\begin{equation}
\left\langle h(t)f^{T}(t')\right\rangle =0,\;t>t'.\label{eq:hfcausality}
\end{equation}
The defining property of $K$ is that the autocorrelation of $u(t)$
reconstructs correctly, so we require $\kB T\left(G(t-t')+G^{T}(t'-t)\right)=\int K(t-t")K^{T}(t'-t")dt".$

These equations have a natural time convolution structure, starting
with 
\begin{equation}
dx/dt=G*f+K*h.
\end{equation}
We will use $R$ to denote negation of time arguments, equivalent
in effect to a transpose of the times as indices. Then we can write
\begin{equation}
\left\langle uu^{T}\right\rangle =\kB T\left(G+G^{TR}\right)=K*K^{TR}.\label{eq:GKreln}
\end{equation}
The above can be interpreted as an L-U matrix decomposition with respect to indices corresponding to times, and we will also have need of a partner U-L decomposition so that 
\begin{equation}
\kB T\left(G+G^{TR}\right)=K*K^{TR}=L^{TR}*L.\label{eq:GKreln2}
\end{equation}
where $L$ is strictly causal.

\subsection{Time Reversal Transformation in terms of Innovation}

If we simply substitute all time arguments in terms of $t^{R}=-t$ in our equation
of motion we obtain
\begin{equation}\label{eq:negatedeom}
-dx^{R}/dt^{R}=G^{R}*f^{R}+K^{R}*H.
\end{equation}
Here $x^{R}(t^{R})=x(t)=x(-t^{R})$ and similarly for $f^{R}$, $G^R$ and $K^R$ as in the memoryless case, and $H(t_R)=h(-t^R)$.  However $G^{R}\equiv G^{R}(t^{R})=G(t)=G(-t^{R})$ and similarly
for $K^{R}$ have the effect that they are propagating backwards in reversed time $t^R$, so the form of the equation of motion is clearly not preserved under this simple time negation. 

The time reversed equation we want is 
\begin{equation}\label{eq:reversedeom}
dx^{R}/dt^{R}=G^T*f^{R}+L^T*h^{R},
\end{equation}
where $G^T$ and $L^T$ are now propagating forwards in their implied arguments of reverse time. The transpose on $G$ here is simply the natural time reverse of the Fluctuation-Dissipation Theorem, and we have chosen our notation for $L$ to match.  This requires 
\begin{equation} \label{eq:defRh}
h^{R}=- (L^T)^{-1}*\left( (G^T+G^{R})*f^{R}+K^{R}*H\right),
\end{equation}
where convolution by $(L^T)^{-1}$ inverts convolution by $L^T$. 

The time reversed partner to Eq.~\eqref{eq:hfcausality} is  $\left\langle h^R(t^{R})f^{TR}({t'}^{R})\right\rangle =0$
for $t^{R}>{t'}^{R}$. Substituting $h_R$ from  Eq.\eqref{eq:defRh} this leads to
\begin{equation}\begin{split}
    (L^T)^{-1}*& K^R*\langle H f^{TR}({t'}^R)\rangle \\&=-(L^T)^{-1}*(G^T+G^R)*\langle f^R f^{TR}({t'}^R)\rangle,
\end{split}
\end{equation}
all for $t^{R}>{t'}^{R}$.  This can then be expressed in terms of unreversed times $t<t'$  as
\begin{equation}\label{eq:hfanticausal}
    (L^{TR})^{-1}*K*\langle h f^{T}({t'})\rangle=-L*\langle f f^{T}({t'})\rangle/(\kB T)
\end{equation}
where on the RHS we have made use of $(L^{TR})^{-1}*(G+G^{TR})=(L^{TR})^{-1}*K*K^R/(\kB T)=L/(\kB T)$ all from Eq. \eqref{eq:GKreln2}.  We can also use \eqref{eq:GKreln2} to substitute $(L^{TR})^{-1}*K= L*(K^{TR})^{-1}$ on the LHS of Eq. \eqref{eq:hfanticausal} to obtain

\begin{equation}\label{eq:hfanticausal2}
    L*(K^{TR})^{-1}*\langle h f^{T}({t'})\rangle=-L*\langle f f^{T}({t'})\rangle/(\kB T)
\end{equation}
for $t<t'$.

\subsection{The cross correlation of $h$ and $f$\label{sect:reversible}}
We now seek to bring equations \eqref{eq:hfcausality} and \eqref{eq:hfanticausal2}
together as a single expression for $\left\langle hf^{T}\right\rangle $. Given that \eqref{eq:hfanticausal2} is valid for $t<t'$ it follows that it remains valid if we convolve both sides forwards to a new time $t$ which is still less than $t'$, and using $L^{-1}*$ in particular then leads to the top line in 
\begin{equation}
(K^{TR})^{-1}*\left\langle hf^{T}\right\rangle (t,t')=-\frac{1}{\kB T}\begin{cases}
\left\langle ff^{T}\right\rangle  & t<t'\\
0 & t>t'
\end{cases}.
\end{equation}
The bottom line of the above follows simply by convolving Eq.~\eqref{eq:hfcausality} by  \((K^{TR})^{-1}\) which being anticausal preserves its time inequality.  Finally it is convenient to denote the braced factor of the RHS above as $\left\langle ff^{T}\right\rangle _{-}$  and then write the whole relation as  
\begin{equation}
    \label{eq:hfcorrn}
\left\langle hf^{T}\right\rangle =-K^{TR}*\left\langle ff^{T}\right\rangle _{-}/(\kB T)
\end{equation}
which can be convolved by $K$ to yield also
\begin{equation}
\left\langle uf^{T}\right\rangle =-(G+G^{TR})*\left\langle ff^{T}\right\rangle _{-}.
\end{equation}

\subsection{Application to correlation function}
Now we come to the heart of the matter and consider 
\begin{equation}
\begin{split}
\left\langle \dot{x}\dot{x}^{T}\right\rangle =&\left\langle uu^T\right\rangle+G*\left\langle ff^{T}\right\rangle *G^{TR}   \\ & +\left\langle uf^{T}\right\rangle *G^{TR}+G*\left\langle fu^{T}\right\rangle 
\end{split}
\end{equation}
where the first term direct from the innovation autocorrelation corresponds
to the ``free'' value of the LHS in the absence of any internal
forces $f$.  In the third term we can substituted \(\left\langle uf^{T}\right\rangle\) using the previous subsection \ref{sect:reversible} to give \begin{equation}
\left\langle uf^{T}\right\rangle *G^{TR} =-(G+G^{TR})*\left\langle ff^{T}\right\rangle _{-}*G^{TR}
\end{equation}
and the fourth term is this value both transposed and time reversed for
which we need to note $\left(\left\langle ff^{T}\right\rangle _{-}\right)^{TR}=\left\langle ff^{T}\right\rangle _{+}$
where $\left\langle ff^{T}\right\rangle =\left\langle ff^{T}\right\rangle _{+}+\left\langle ff^{T}\right\rangle _{-}$,
leading net of some cancellations to
\begin{equation}
\begin{split}
\AF(t-t') &= \left\langle \dot{x}\dot{x}^{T}\right\rangle _{\textrm{free}}-\left\langle \dot{x}\dot{x}^{T}\right\rangle \\& = G^{TR}*\left\langle ff^{T}\right\rangle _{-}*G^{TR}+G*\left\langle ff^{T}\right\rangle _{+}*G.
\end{split}
\end{equation}

The instantaneous case introduced by Fixman now serves as a check.  For $G(t-t')=M\delta(t-t')$,
we have $G=G^{R}$ and the above result simplifies down to Fixman's Law, 
\begin{equation}
\left\langle \dot{x}\dot{x}^{T}\right\rangle _{\textrm{free}}-\left\langle \dot{x}\dot{x}^{T}\right\rangle =\left\langle vv^{T}\right\rangle 
\end{equation}
where $v(t)=M\,f(x(t))$ is the direct contribution to the motion
due to the conservative forces.

\section{Worked Example of Fixman's Law: Ornstein-Uhlenbeck process in memory medium}

Our Langevin equation \eqref{eq:memorylangevin} becomes linear and directly solvable by Fourier methods, if we specialise to the case where the conservative forces are linear in the displacement coordinates, 
\begin{equation}
    f(x(t))=-s.x(t),
\end{equation}
where $s$ is a real symmetric matrix with non-negative eigenvalues.  This solvable Ornstein-Uhlenbeck process \cite{OUprocess} then provides a direct check on the full memory medium version of Fixman's law \eqref{eq:AFfixmanMM}.

The Fourier Transform of the Langevin equation is now 
\begin{equation}\label{eq:fourierlangevin}
    (i\omega + G(\omega)s)x(\omega)=u(\omega)
\end{equation}
where $\langle u(\omega)u^+(\omega) \rangle = 2 \kB T \left(G(\omega)+G^+(\omega)\right)$ and $G^+={G^*}^T$ is the Hermitian conjugate, in which complex conjugation takes over the role of reversing time.  It then follows directly that 
\begin{equation}\begin{split}
    \langle x&(\omega)x^+(\omega) \rangle = \kB T (i\omega +Gs)^{-1}(G+G^+)((i\omega +Gs)^T)^{-1}\\
    &=\kB T \left( (i\omega +G s)^{-1}s^{-1} + \textrm{Hermitian conjugate} \right).
\end{split} \end{equation}

We can immediately deduce the force correlations and noting that $G$ and hence $i\omega+G$ is causal we identify
\begin{equation}
\langle f(\omega)f(\omega)^+ \rangle_+ = \kB T s(i\omega +G s)^{-1}
\end{equation}
with $\langle f(\omega)f^+(\omega) \rangle_-$ given by the Hermitian conjugate and the subscripts $\pm$ relate back to the time domain versions.

The  analogously calculation for $\langle \dot{x}(\omega)\dot{x}^+(\omega) \rangle_+$ is slightly more involved.  We start from 
\begin{equation}\begin{split}
\langle \dot{x}(\omega)\dot{x}(\omega)^+ \rangle =& \kB T \left( G - i\omega s^{-1} - Gs(i\omega+Gs)^{-1} G \right)  \\ +& \textrm{ Hermitian conjugate } .
\end{split} \end{equation}
Then after cancelling  terms in $\mp i\omega s^{-1}$ we can read off that
\begin{equation}
    \langle \dot{x}(\omega)\dot{x}(\omega)^+ \rangle_+=\langle u(\omega)u^+(\omega) \rangle_+ -  G\langle f(\omega)f(\omega)^+ \rangle_+ G
\end{equation}
in full agreement with Fixman's Law extended to a memory medium \eqref{eq:AFfixmanMM}.
\end{document}